\documentstyle[epsfig,12pt]{article}

\begin{document}

\hoffset = -1truecm \voffset = -2truecm \baselineskip = 10 mm

\title{\bf Can a chaotic solution in the QCD evolution equation restrain
    high-energy collider physics?
}

\author{
 Wei Zhu,  Zhenqi Shen and Jianhong Ruan \\\\
\normalsize Department of Physics, East China Normal University,
Shanghai 200062, P.R. China}

\date{}

\newpage

\maketitle

\vskip 3truecm

\begin{abstract}

    We indicate that the random aperiodic oscillation of the gluon
distributions in a modified Balisky--Fadin--Kuraev--Lipatov (BFKL)
equation has positive Lyapunov exponents. This first example of
chaos in QCD evolution equations, raises the sudden disappearance of
the gluon distributions at a critical small value of the Bjorken
variable $x$ and may stop the increase of the new particle events in
an ultra high energy hadron collider.

\end{abstract}


{\bf keywords}:  QCD evolution equation; Chaos

\newpage

    The gluon distributions are important knowledge in research of
high energy collider physics. The gluon distributions of the nucleon
cannot be extracted directly from the measured structure functions
in deep inelastic scattering experiments and they are mainly
predicted by using the QCD evolution equations. However, the linear
Dokshitzer-Gribov-Lipatov-Altarelli-Parisi (DGLAP)$^{1,2}$ and
Balitsky-Fadin-Kuraev-Lipatov (BFKL)$^3$ evolution equations are no
longer reliable at ultra higher energy, therefore a series nonlinear
evolution equations, for example, the GLR-MQ-ZRS
Gribov-Levin-Ryskin, Mueller-Qiu, Zhu-Ruan-Shen (GLR-MQ-ZRS)$^{4-6}$
and Balitsky-Kovchegov (BK)$^7$ equations were proposed, in which
the corrections of the gluon fusion are considered. An important
prediction of these equations is that the gluon distributions
approach to a so-called `saturation limit' asymptotically at small
Bjorken variable $x$, where gluon splitting is balanced with fusion.

    As we well know, the nonlinear dynamics may have a characteristic solution--chaos,
which have been observed in many natural phenomena$^8$. Therefore,
it is interesting to ask: Does nonlinear QCD evolution equation of
gluon distribution also have chaotic solution? How does it impact
the gluon distribution? Recently, we have proposed a nonlinear
modified BFKL (MD-BFKL) equation in Ref. [9], which describes the
corrections of the gluon recombination to the BFKL equation. We
found that the unintegrated gluon distribution function $F(x, k^2)$
in the MD-BFKL equation in the fixed coupling approximation begins
its smooth evolution under suppression of the gluon fusion likes the
solution of the BK equation, but when $x$ comes to a smaller
critical $x_c$, $F(x, k^2)$ will oscillate aperiodically near the
evolution endpoint $k^2_0$. This behavior has the characteristic
feature of chaos: random and sensitive to the initial conditions.
Furthermore, with the enhancement of oscillation, the distribution
$F(x, k^2)$ will disappear suddenly at $x_c$. This kind of phase
transition led by chaotic solutions will certainly call our
attention to reconsider the future of high energy collider physics
carefully.

    In this letter, we indicate that the above-mentioned oscillate solution in the MD-BFKL equation is really
chaos and it still holds in the running coupling case.

\vskip -3.0 truecm
 \hbox{

\centerline{\epsfig{file=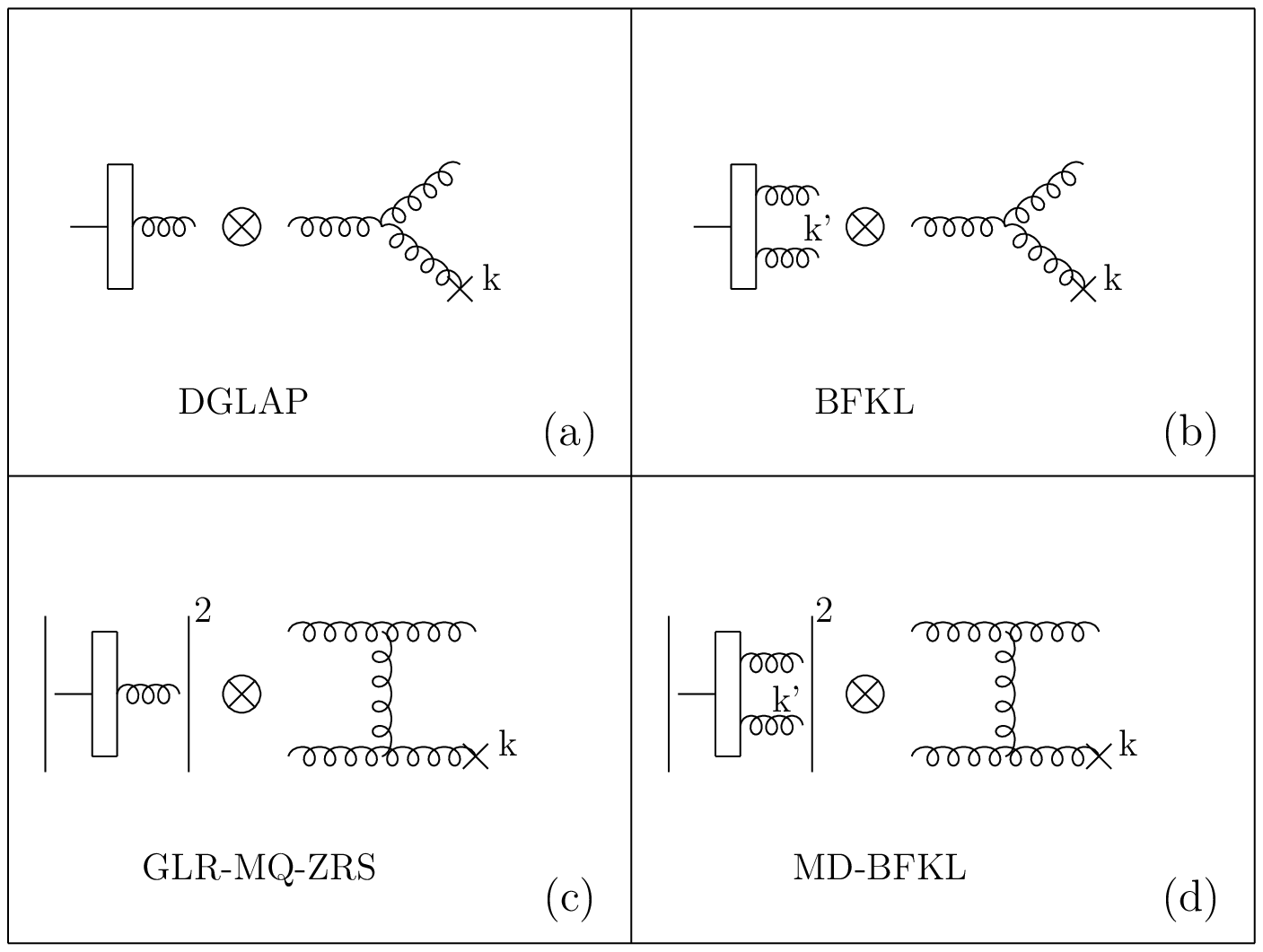,width=15.0cm,clip=}}}

\vskip -9.0 truecm

 Fig. 1: A part of the amplitudes for four related
evolution equations. (a) The DGLAP equation; (b) the BFKL equation;
(c and d) the GLR-MQ-ZRS and MD-BFKL equations, which contain the
leading corrections of the gluon recombination to the DGLAP and BFKL
equations, respectively.

\vskip 1.0 truecm

    We develop the MD-BFKL equation$^9$ due to the following considerations: as a
leading logarithmic approximation, the DGLAP equation neglects the
correlation of initial partons. With the increase of parton
densities, more initial gluons should be considered in evolution. By
adding initial gluons on the elementary amplitude of the DGLAP
equation in Fig. 1(a) step by step, we can reach the amplitudes in
Figs. 1(b)-1(d). In a unified theoretical framework and by making
use of time ordered perturbation theory (TOPT),$^{10}$, we derived
the well-known the BFKL and GLR-MQ-ZRS equation as well as a new
MD-BFKL equation, which reads

    $$-x\frac {\partial F(x,k)}{\partial x}$$
$$=\frac{\alpha_{s}N_c}{\pi^2}\int d^2k'
\frac{k^2}{k'^2(k'-k)^2}\left[F(x,k')-\frac{1}{2}F(x,k)\right]$$
$$+\frac{18\alpha^2_s}{\pi^2R^2_N}\frac{N_c^2}{N_c^2-1}
\int d^2k' \frac{k^2}{k'^2(k'-k)^2}\left[\frac{1}{k'^2}
F^2\left(\frac{x}{2},k'\right)-\frac{1}{2k^2}F^2\left(\frac{x}{2},k\right)\right]$$
$$-\frac{36\alpha^2_s}{\pi^2R^2_N}\frac{N_c^2}{N_c^2-1}
\int d^2k' \frac{k^2}{k'^2(k'-k)^2}\left[\frac{1}{k'^2}
F^2(x,k')-\frac{1}{2k^2}F^2(x,k)\right], \eqno(1)$$ where the three
terms on the right-hand side are the BFKL kernel of gluon splitting,
nonlinear anti-shadowing and shadowing kernels led by gluon fusion,
respectively. Each term includes real and virtual parts in order to
ensure that the evolution kernels are infrared safe. It should be
pointed out that Eq. (1) is different from the BK equation both in
their elementary amplitudes and structures. Although the derivation
of this equation is tedious$^9$, the reason of it can be proven by
the consistency among this equation and DGLAP, BFKL and GLR
equations. In fact, as we see in Fig. 1, the structures of the
MD-BFKL and GLR-MQ-ZRS equations have the similar relationship of
those in the BFKL and DGLAP equations: at leading approximation, the
elementary amplitudes of the BFKL and DGLAP equations share the same
evolution kernel-gluon splitting. However, in the BFKL-initial
state, there are two gluons correlated by relative transverse
momentum $k^{\prime}$, therefore the initial state in the BFKL
equation can connect with the gluon splitting vertex by two
different ways. With simple algebra, we can find that the factor
$1/k^2$ in the DGLAP evolution kernel becomes
$k^2/k^{{\prime}2}(k'-k)^2$ in the BFKL equation because of the
contributions of the interference diagrams. Similarly, the
GLR-MQ-ZRS and MD-BFKL equations have the same gluon recombination
kernel, but in the later equation the evolution kernel connect with
the initial state also by two ways. Replacing the evolution kernel
in the GLR-MQ-ZRS equation in such ways, one can reach the MD-BFKL
equation. That is, once the DGLAP, BFKL and GLR-MQ-ZRS equations are
determined, the form of the MD-BFKL equation (1) is also fixed. We
should emphasize that a complete evolution equation must include
contributions from virtual diagrams for infrared safety, however,
they are cancelled in the GLR-MQ-ZRS equation, while cannot be
neglected in the MD-BFKL equation.

    In this work, we use a running coupling $\alpha_s(k^2)$ in Eq. (1)
and so we need to consider the diffusion in $\ln k^2$ with
decreasing $x$, which leads to an increasingly large contribution
from the infrared region of $k^2$ where the equation is not expected
to be valid. For this sake, as Ref. [11] we split the integration
region for real gluon emission in Eq. (1) up into two parts:
region(A) 0 to $k^2_0$ and region(B) $k^2_0$ to $\infty$. In
region(B) the MD-BFKL equation as it stands is taken to hold and in
region(A) $F(x,k^2)$ is parameterized as $C k^2/(k^2+k^2_a)$ with
$k^2_a=1$ GeV$^2$, where the parameter $C$ keeps the smooth
connection between two parts. For simplicity, we neglect the
contributions from the antishadowing effects and take the
cylindrically symmetric solution. We use the Runge-Kutta method to
compute the evolution equations. Assuming that a symmetric Gaussian
input distribution exits at the evolution stating point
$x_0=10^{-3}$.

$$F(x_0=10^{-3},k^2)(k^2)^{-\frac{1}{2}}=\beta
exp\left[-\frac{\log^2 (k^2/1 GeV^2))}{40}\right], \eqno(2)$$ where
$\beta=0.1$. The solutions of equation (1) are shown in Fig.2.
Firstly, we can see that the distribution $F(x, k^2)$ disappear
suddenly at $x=x_c$. Furthermore, curves with different values of
$k^2$ turn down at the same $x_c$. If we regard $x$ as an order
parameter, Fig. 2 exhibits a phase transition of first kind.

\hbox{
 \centerline{\epsfig{file=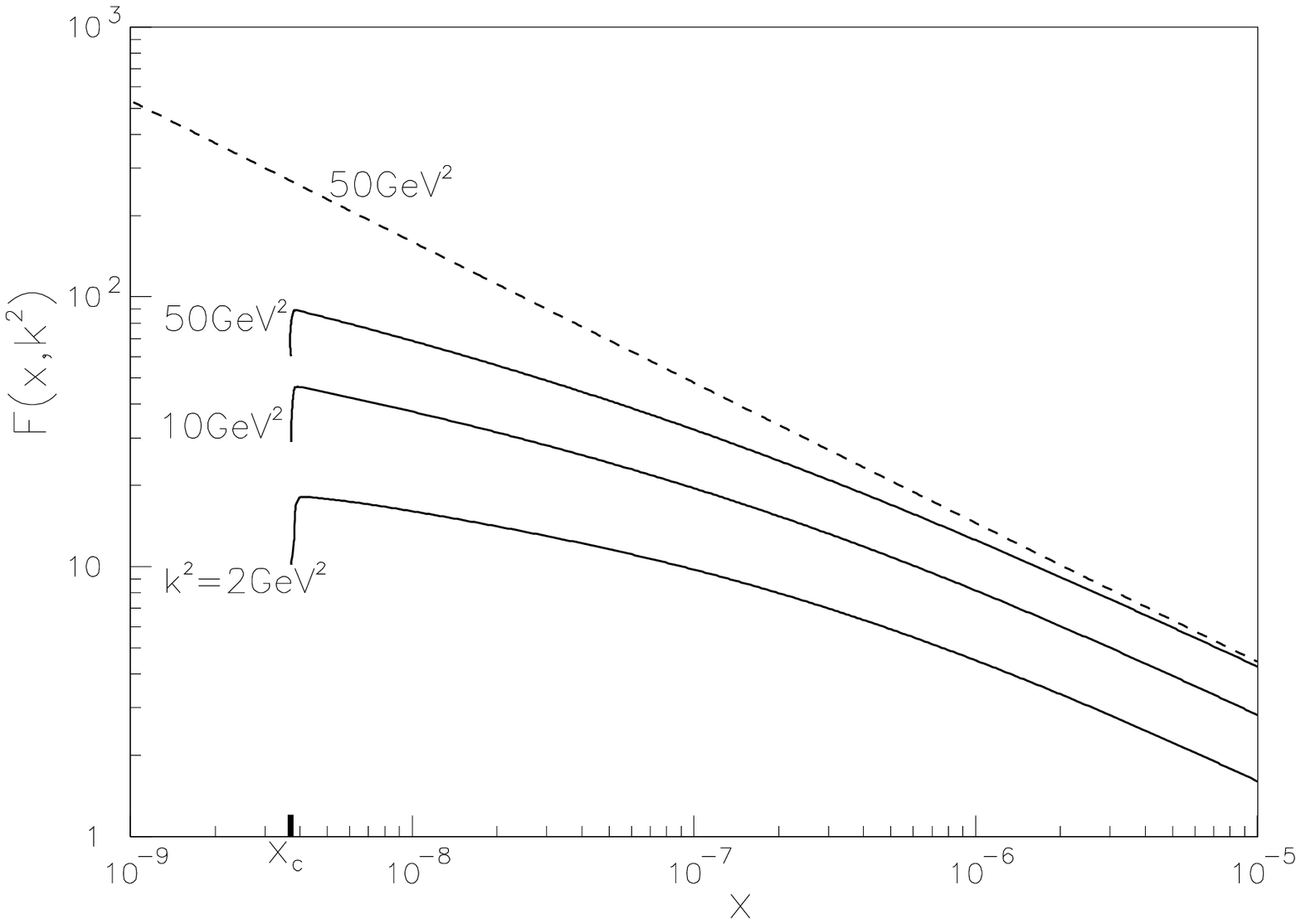,width=15.0cm,clip=}}}

\vskip -1.0 truecm Fig.2:  Dependence of the unintegrated gluon
distribution on $x$ in the MD-BFKL equation (1) for different values
of $k^2$. The results show that $F(x,k^2)$ suddenly drops near a
critical value of $x\sim x_c\simeq 3.8\times 10^{-8}$. The dashed
curves are the corresponding solution of the BFKL equation with
$k^2=50GeV^2$.

\vskip 1.0 truecm

    The reason of the sudden disappearance of the gluon distribution can
be innovated by the relation of $F(x, k^2)$ and $k^2$ in Fig.3.
From the start point $x_0$, gluons diffuse on the transverse
momentum space rapidly under the action of the BFKL linear kernel.
Because of the gluon fusion, the above diffusion towards low $k^2$
is suppressed obviously. All these are like that the BK equation
has predicted. While what is different is that, when $x$ goes to
$x_c$, the oscillation of the curve will happen and increase
rapidly once it near $k^2_0$. This leads to a huge shadowing
effect and the distribution $F(x, k^2)$ disappears at $x_c$. It is
interesting that dispersed gluons will gather near $k^2_0$ before
disappearance. This kind of oscillation is random because $k$ is
not ordered in evolution. Furthermore, the above aperiodic
oscillation is very sensitive to the initial conditions.
Especially, the oscillation will be enhanced with the increase of
the numerical calculating precision. These features are also
observed in other chaos phenomena universally.

\hbox{
 \centerline{\epsfig{file=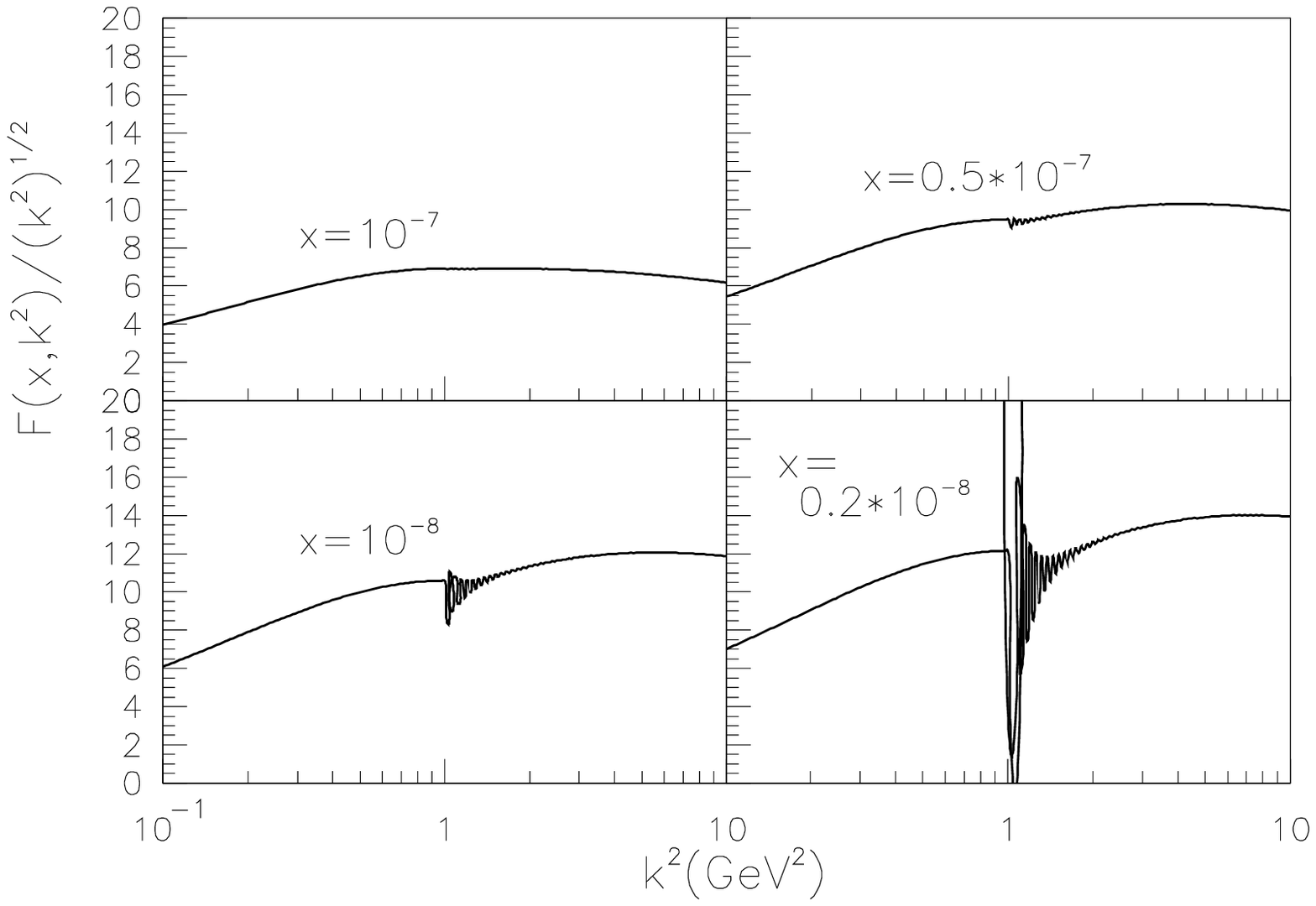,width=15.0cm,clip=}}}
\vskip -1.0 truecm

 Fig.3: Dependence of the unintegrated gluon
distribution on $k^2$ for different values of $x$. The results
present the oscillations near $x_c$ when $k^2\rightarrow k^2_0$.
This leads to a huge shadowing effect and the distribution $F(x,
k^2)$ disappears at $x<x_c$.

\vskip 1.0 truecm

   A standard criterion of chaos is that the system has the
positive Lyapunov exponents, which indicates a strong sensitivity to
small changes in the initial conditions$^8$. We regard $y=\ln 1/x$
as `time' and calculate the Lyapunov exponents $\lambda(k^2)$ in a
finite region: $10^{-7}\le x\le 0.2\times 10^{-8}$, where the
oscillation of the distribution is obvious. We divide equally the
above mentioned $y$-region into n parts with $y_1,y_2.., y_{n+1}$
and $\tau=(y_{n+1}-y_1)/n$. Assuming that the distribution evolves
to $y_1$ from $y_0=\ln 1/x_0$ and results $F(y_1,k)$. Corresponding
to a given value $F(y_1,k)$ at $(y_1,k)$, we perturb it to
$F(y_1,k)+\Delta$ with $\Delta\ll 1$. Then we continue the
evolutions from $F(y_1,k)$ and $F(y_1,k)+\Delta$ to $y_2$ from $y_1$
respectively, and denote the resulting distributions as $F(y_2,k)$
and $\tilde{F}(y_2,k)$. Making the difference
$\Delta_2=\vert\tilde{F}(y_2,k)-F(y_2,k)\vert$. In the following
step, we repeat the perturbation $F(y_2,k)\rightarrow
F(y_2,k)+\Delta$ and let the next evolutions from $F(y_2,k)$ and
$F(y_2,k)+\Delta$ from $y_2$ to $y_3$ respectively and get the
results $\Delta_3= \vert\tilde{F}(y_3,k)-F(y_3,k)\vert$...... (see
Fig. 4). The Lyapunov exponents for the image from $y$ to $F(y,k)$
are defined as

$$\lambda(k^2)=\lim_{n\rightarrow\infty}\frac{1}{n\tau}\sum_{i=2}^{n+1}\ln\frac {\Delta_i}
{\Delta}. \eqno(3)$$ The Lyapunov exponent of the gluon
distribution in the MD-BFKL equation with the input Eq. (2) are
presented in Fig. 5. For comparison, we give the Lyapunov
exponents but using the BFKL and BK equations in Fig. 5. The
positive values of the Lyapunov exponents clearly show that the
oscillation of $F(x, k){\sim}k^2$ is chaos of the MD-BFKL
equation. Therefore, we conclude that chaos in the MD-BFKL
equation lead to the sudden disappearance of the gluon
distributions.

\vskip -8.0 truecm
 \hbox{
 \centerline{\epsfig{file=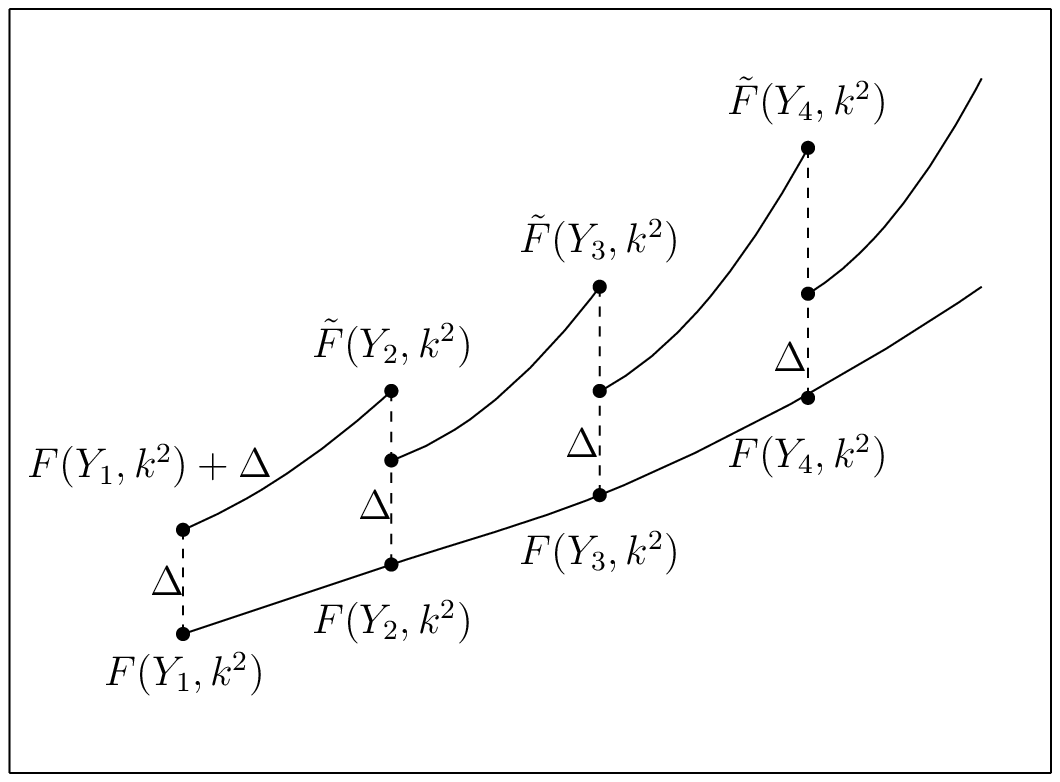,width=15.0cm,clip=}}}
\vskip -12.0 truecm

 Fig. 4: Schematic programs calculating the
Lyapunov~exponents~of~the~evolution~equation. The results are
insensitive to the value of $\Delta$.

\vskip 1.0 truecm
 \hbox{
 \centerline{\epsfig{file=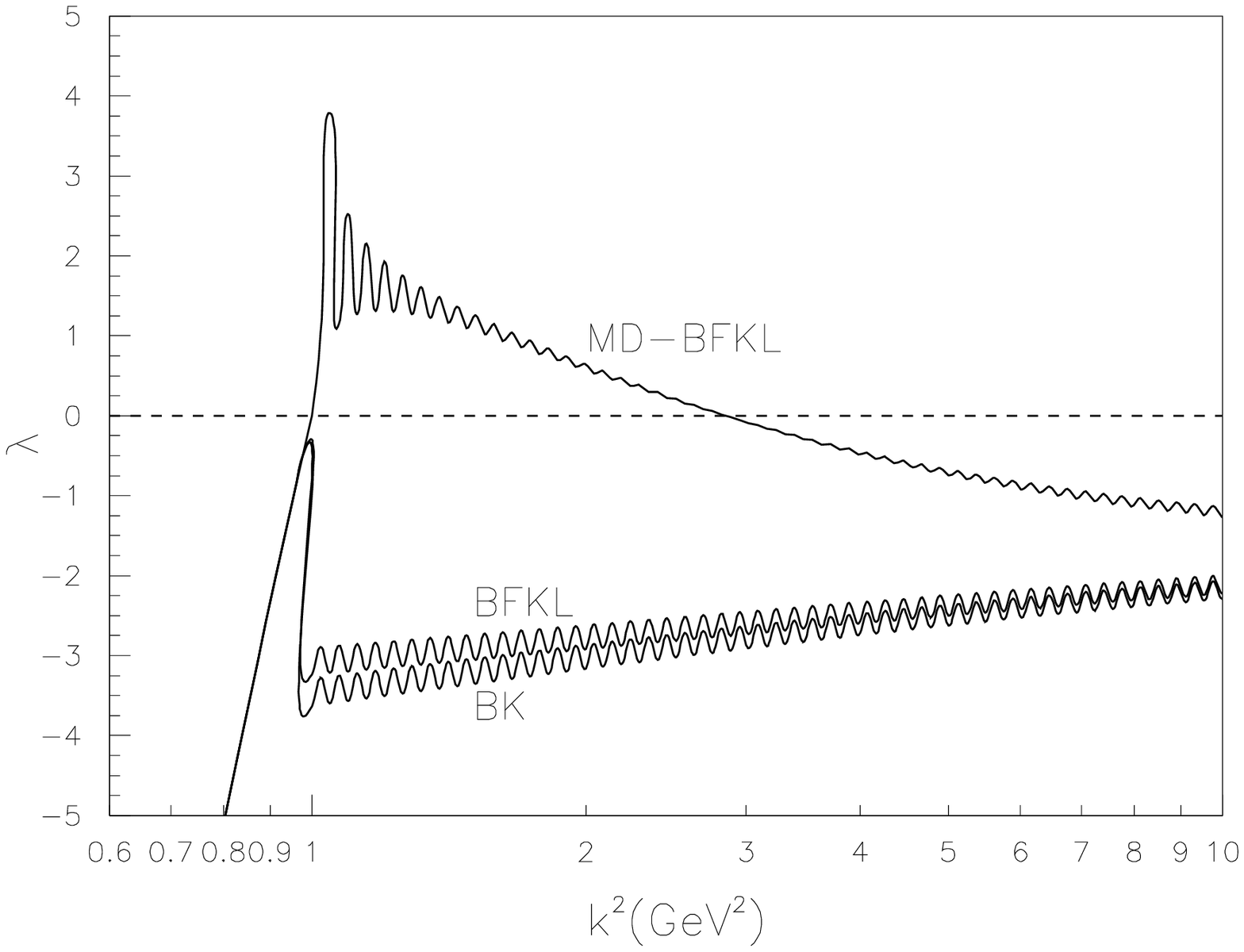,width=15.0cm,clip=}}}
Fig. 5: Plots of the Lyapunov exponents in the region $10^{-7}\le
x\le 0.2\times 10^{-8}$. The positive Lyapunov exponents show that
the corresponding solution of the MD-BFKL equation is chaos.

\vskip 1.0 truecm

    The important questions are at which scale gluons will
disappear and how much it will impact negatively on high energy
collider physics. Unfortunately, we cannot predict the start point
of the evolution of MD-BFKL equation at this moment. But considering
that the MD-BFKL equation works right after GLR-MQ-ZRS and BK
equations, gluon disappearance should happen after the saturation
phenomena predicted by GLR-MQ-ZRS and BK equations.

    A typical process testing new particle with mass M on high energy hadron
collider, for example in the gluon fusion model it directly
relates to the unintegrated gluon distribution via

$$\sigma=\int d^2k_1\int d^2k_2\int_0^1dx_1\int_0^1dx_2\delta(x_1x_2-\tau)\delta^{(2)}(k_1+k_2-p_T)
F_1(x_1,k_1)F_2(x_2,k_2)d\hat{\sigma}, \eqno(4)$$ which is the
function of $\tau=M^2/s$. The increase of the new particle events
with increasing energy $s$ will be stopped due to the gluon
disappearance when $\tau\le x_c$. Thus we need to re-estimate the
trials to the new physics in hadron collider physics.

    Finally, we emphasize that the MD-BFKL equation
is constructed based on a naive partonic picture (Fig. 1) and simple
leading QCD corrections. Many higher order corrections are
neglected, such as possible mixture of the operators with different
twists, the next leading logarithmic (NLL)) and next leading order
(NLO) corrections, singularities from non-perturbative parts in the
factorization procedure. Of course, the MD-BFKL equation will only
be an applicable QCD evolution equation until all the above
corrections are considered. An important questions is: will chaos
effect we demonstrated in the MD-BFKL equation disappear after
further corrections are considered? To answer this question, we
point out that chaos does not appear in the nonlinear GLR equation
is related to the fact that its evolution kernel has no singularity.
On the other hand, although the evolution kernel in the nonlinear
term of the BK equation is also the singular BFKL kernel, these
singularities can be absorbed into the re-definition of the
scattering amplitudes, like its form in momentum space. Therefore
the BK equation has no chaotic solution, either. Thus we suggest the
fact that chaos appear in the MD-BFKL equation firstly is related to
the singularities in its nonlinear terms. From the experiences in
the study of the BFKL kernel, those possible QCD corrections we
mentioned to the MD-BFKL will probably make singularities of this
nonlinear equation more complicated, instead of removing these
singularities completely, since the regularized singular parts of
the evolution kernel always dominate the evolution. In this
situation, we could expect that more interesting chaos phenomena
will appear in the new MD-BFKL equation. These phenomena will be
most interesting to high energy physics and nonlinear science.

    In summary, a random oscillation of the unintegrated gluon
distributions in a modified BFKL equation is indicated as chaos,
which has the positive Lyapunov exponents. This first example of
chaos in QCD evolution equations, raises the sudden disappearance
of the gluon distributions at a critical small value of the
Bjorken variable $x$ and may stop the increase of the new particle
events in a ultra high energy hadron collider.

\vspace{0.3cm}

\noindent {\bf Acknowledgments}:  We thank Z.H. Liu for useful
discussions in chaos. This work was supported by National Natural
Science Foundations of China 10475028.

\end{document}